\documentclass[superscriptaddress,showpacs,showkeys,twocolumn,prb,aps,reprint,a4paper,floatfix]{revtex4-2}
\usepackage[english]{babel}
\usepackage[utf8]{inputenc}
\usepackage[colorinlistoftodos, color=green!40, prependcaption]{todonotes}
\usepackage[pdftex, pdftitle={Article}, pdfauthor={Author}]{hyperref} 

\usepackage{xcolor}

\usepackage{amsmath}
\usepackage{amsfonts}

\usepackage{graphicx}
\usepackage{subcaption}
\usepackage{float}
\usepackage{caption}

\usepackage{tikz}
\usetikzlibrary{shapes.geometric, arrows, positioning}
\tikzstyle{startstop} = [rectangle, rounded corners, minimum width=2cm, minimum height=1cm, text centered, draw=black, fill=red!30]
\tikzstyle{process} = [rectangle, minimum width=2cm, minimum height=1cm, text centered, draw=black, fill=orange!30]
\tikzstyle{thermal} = [rectangle, minimum width=2cm, minimum height=1cm, text centered, draw=black, fill=blue!30]
\tikzstyle{arrow} = [thick,->,>=stealth]

\begin{document}

\graphicspath{{Figs/}}

\title{Superconducting Ring Resonators: Modelling, Simulation, and Experimental Characterisation}

\author{Zhenyuan Sun}
    \email{zs311@cantab.ac.uk}
    \affiliation{Cavendish Laboratory, JJ Thomson Avenue, Cambridge, CB3 0HE, United Kingdom.}
\author{S Withington}
   \affiliation{Clarendon Laboratory, Parks Road, Oxford, OX1 3PU, United Kingdom.}
\author{C N Thomas}
    \affiliation{Cavendish Laboratory, JJ Thomson Avenue, Cambridge, CB3 0HE, United Kingdom.}
\author{Songyuan Zhao}
   \affiliation{Clarendon Laboratory, Parks Road, Oxford, OX1 3PU, United Kingdom.}

\date{June 30, 2025}

\begin{abstract}
We present a theoretical and experimental study of superconducting ring resonators as an initial step toward their implementation in superconducting electronics and quantum technologies, with promising applications including superconducting parametric amplifiers with pump-signal isolation, flux-controlled quantum circuits, ultra-sensitive measurements in quantum sensing, and THz instrumentations. These devices have the potentially valuable property of supporting two orthogonal electromagnetic modes that couple to a common Cooper pair, quasiparticle, and phonon system. We present here a comprehensive theoretical and experimental analysis of the superconducting ring resonator system.
We have developed superconducting ring resonator models that describe the key features of microwave behaviour to first order, providing insights into how transmission line inhomogeneities give rise to frequency splitting and mode rotation.    
Furthermore, we constructed signal flow graphs for a four-port ring resonator to numerically validate the behaviour predicted by our theoretical analysis. Superconducting ring resonators were fabricated in both coplanar waveguide and microstrip geometries using Al and Nb thin films. Microwave characterisation of these devices demonstrates close agreement with theoretical predictions. Our study reveals that frequency splitting and mode rotation are prevalent in ring systems with coupled degenerate modes, and these phenomena become distinctly resolved in high quality factor superconducting ring resonators.
\end{abstract}

\keywords{superconducting resonator, ring resonator, resonator characterisation, coupled mode theory, signal flow graph, frequency splitting}

\maketitle

\section{Introduction}\label{sec_introduction}
Superconducting thin-film resonators are being developed for a wide range of applications in low temperature physics, for example, as sensitive detectors, especially in the context of kinetic inductance detectors (KIDs) \cite{day2003broadband, zmuidzinas2012superconducting}, for mm/sub-mm \cite{day2006antenna, schlaerth2008millimeter, endo2016superconducting}, near-infrared and optical \cite{o2011arcons, gao2012titanium} astrophysics, and also for particle physics, such as x-ray imaging \cite{vardulakis2007superconductingastrophysics}, dark matter searches \cite{daal2008kinetic, golwala2008wimp}, and neutrino mass measurements \cite{faverzani2012developments, di2014cryogenic,QTNM_collaboration_white_paper}.  In addition to their direct application as sensors, superconducting resonators also demonstrate versatility in other important roles, as  ultra-low-noise superconducting parametric amplifiers \cite{Tholen_2009_paper,Zhao_2023,zhao2024_intrinsic_separation} and as critical elements in superconducting quantum interference devices (SQUIDs) frequency domain multiplexing for transition edge sensors (TESs) \cite{mates2008demonstration}. In quantum computing, superconducting resonators have become key components for qubit control and readout \cite{wallraff2004strong, schoelkopf2008wiring, sillanpaa2007coherent}. More generally, superconducting resonators have also been used to couple quantum devices for exploring chip-based quantum electrodynamics, such as nano-mechanical resonators \cite{hertzberg2010back} and the spin state of a single-molecule magnet \cite{fan2011measuring}.

Most physical implementations of superconducting resonators are based on microstrip transmission lines, coplanar waveguides (CPW), or lumped element circuits \cite{zmuidzinas2012superconducting, mauskopf2018transition, mcrae2020materials}. Superconducting films (usually elemental metals or their alloys, e.g. Al, Nb, Ti, NbN, NbTiN) are deposited on dielectric substrates (e.g.  Si, SiN, sapphire) and patterned using etching or lift-off processes in combination with photolithography. These superconducting films, typically described by BCS and Ginzburg-Landau theories, exhibit properties that make them ideally suited for advanced technologies, including strong reactive nonlinearity for quantum-limited parametric amplifiers \cite{Eom_2012,zhao2022physics}, low dissipation for high-Q resonators \cite{Barends_2008,Megrant_2012}, and high pair-breaking frequency gaps for millimetre-wave and THz instrumentations \cite{Jonas_1998,Kroug_2001}. Superconducting resonators can also be implemented in alternative geometries, such as a closed ring of transmission line, i.e. a superconducting ring resonator. Ring resonators are widely used in both the microwave and photonics communities. In microwave research, developments have primarily focused on ring resonators made from normal metals, with recent work exploring the effects of curvature on dispersion \cite{wolff1971microstrip}, applications at millimetre-wave frequencies \cite{hopkins2008millimetre}, and implementations as passive components such as filters, couplers, splitters, and antennas, as well as active components such as mixers \cite{makimoto2013microwave,chang2004microwave}. In photonics, ring resonators support a broad range of passive applications, including filters, delay lines, and add-drop filters \cite{Rabus2007}, along with active applications such as tunable splitters, modulators, and switches \cite{ring_splitter,Chremmos_2010}. They have also been employed in cavity quantum electrodynamics \cite{Renner_2008}. However, to the best of our knowledge, in-depth analysis of ring resonators based on superconducting transmission lines appears to be limited in the existing literature. As will be demonstrated in this study, compared to conventional resonators with linear geometries, superconducting ring resonators show distinctive properties that enable promising applications: (1) they support ideally degenerate modes with high isolation, which can be used for pump–signal isolation in resonator parametric amplifiers; (2) their intrinsic flux sensitivity makes them well suited to applications in flux-controlled quantum circuits; (3) their pronounced sensitivity to asymmetries and non-uniformities, arising from defects or external fields, leads to frequency splitting that can enable ultra-sensitive differential measurements in quantum sensing. In this manuscript, we define the term `superconducting ring resonator' as a low-temperature-operated superconducting transmission line resonator patterned in a loop geometry \cite{Koch2011ring, ricci2006single, huang2021superconducting}. Our aim is to investigate whether superconducting ring resonators can be realised in practice, and to study their potential uses and issues associated with their design and operation.

Translating the unique properties and advantages of the ring resonator technology from the fields of microwaves \cite{chang2004microwave, hopkins2008millimetre, wolff1971microstrip,chorey1991ybco, makimoto2013microwave} and photonics \cite{Chremmos_2010, jazayerifar2017feasibility, melloni2010tunable, ye2010power} to the field of low temperature superconductivity would lead to impactful research. Superconducting ring resonators offer significant advantages for practical instrumentation: (1) they support two lowest-order ideally degenerate modes within a shared ring structure, allowing independent control of each mode through appropriate coupling schemes, e.g. the four-port ring proposed in this paper; (2) owing to their closed-loop geometry, these resonators have no end effects and exhibit low radiation loss; (3) their high quality factor (high-Q) and low intrinsic loss make them well-suited for low-noise and high-sensitivity applications; (4) they offer many promising potential applications as low-loss superconducting passive components, such as filters, delay lines, couplers, splitters and isolators; (5) their intrinsic response is influenced by both dissipative and reactive nonlinearities, which may be advantageous depending on the context, e.g. the frequency-dependent kinetic inductance is valuable for realising parametric devices. In particular, as shown in this study, a four-port ring resonator supports two lowest-order spatially orthogonal electromagnetic modes that only couple the ports on opposite sides of the ring. These modes can be excited independently of each other and in principle share the same resonance frequency. This unique property of four-port ring resonators makes them highly suitable for nonlinear superconducting electronics, such as superconducting parametric amplifiers, where effective separation of the bias pump tone and signal tone is crucial \cite{zhao2024_intrinsic_separation}.  

\begin{figure}[htbp!]
\includegraphics[width = 0.9\linewidth]{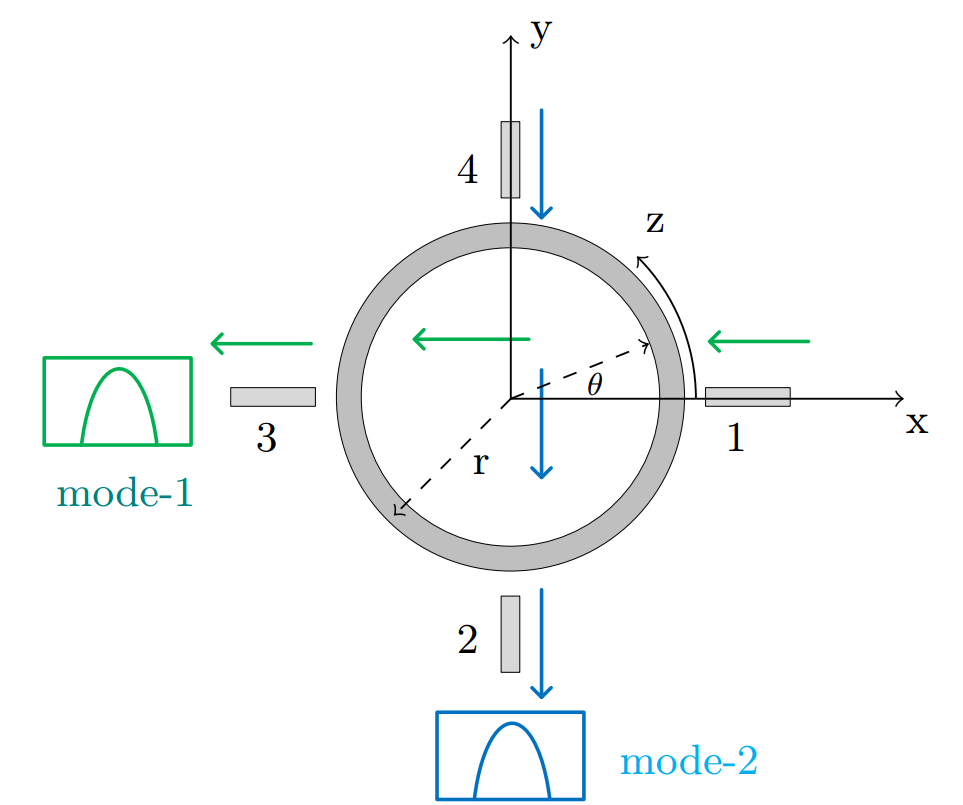}
\caption{\label{fig:221212_superconducting_ring_schematic} Schematic diagram of a four-port superconducting ring resonator. A circular superconducting transmission line of radius $r$ is capacitively coupled to external transmission lines at four evenly spaced points. Two lowest-order orthogonal electrical modes share a common ring system. }
\end{figure}

We present in this paper a comprehensive theoretical and experimental analysis of the superconducting ring resonator system.
We have developed a perturbation analysis of ring resonators to describe the key features of its microwave behaviour to the first order, providing insights into how transmission line inhomogeneities give rise to frequency splitting and mode rotation.    
We constructed signal flow graphs for a four-port ring resonator to numerically validate the predictions from our perturbation analysis \cite{withington2013elastic, guruswamy2018nonequilibrium}. Building on these theoretical foundations, we have designed, fabricated, and characterised a set of superconducting ring resonators based on circular transmission lines of radius $r$ capacitively coupled to the ends of four external readout transmission lines spaced equally around the circumferences, as shown in Fig.\ref{fig:221212_superconducting_ring_schematic}. These devices were realised in both coplanar waveguide and microstrip geometries using superconducting Al and Nb thin films. Microwave characterisation of these devices demonstrates close agreement with theoretical predictions. Our study reveals that frequency splitting and mode rotation are prevalent phenomena in ring systems with coupled degenerate modes, which becomes distinctly resolved in high quality factor superconducting ring resonators.

\section{Theory and Simulations}\label{sec_S2_Superconducting_ring_resoantor_model}
\subsection{MODEL OF RING RESONATOR}\label{subsec_S2_model_of_ring_resonator}
We model the ring as a length of transmission line with series resistance $R$, series inductance $L$, shunt conductance $G$, and capacitance $C$, respectively, per unit length of line \cite{pozar2011microwave}. The voltage $V(z, t)$ and current $I(z, t)$ around the ring are described by Telegrapher's equations 
\begin{equation}\label{eqn:S2_telegrapher_equation}
\frac{\partial V}{\partial z} = - L \frac{\partial I}{\partial t} - RI, \,\,\,\,\, \frac{\partial I}{\partial z} = - C \frac{\partial V}{\partial t} - GV,
\end{equation}
which satisfy the periodic boundary conditions $V(z + 2 \pi r, t)=V(z, t), \, I(z + 2 \pi r, t)=I(z, t)$. Considering the lowest-order modes of free oscillation, the general solution can be approximated by
\begin{equation}\label{eqn:S2_general_solution_of_telegrapherequation_decompostion_1st_order}
\begin{split}
& V(z, t) = V_c(t) \cos \frac{z}{r} + V_s(t) \sin \frac{z}{r},\\
& I(z, t) = I_c(t) \cos \frac{z}{r} +I_s(t) \sin \frac{z}{r},
\end{split}
\end{equation}
where $V_c,\,\, V_s,\,\, I_c,\,\, I_s$ are voltage and current decomposition coefficients. 

Consider an isolated and near-lossless ring, i.e. $\omega L \gg R$ and $\omega C \gg G$, which is sufficient to capture the key behaviour of the device. In order to study the effect of inhomogeneities along the superconducting transmission lines, we perform perturbation analysis by expanding in terms of the perturbed quantities $L(z) = L_0 + \delta L(z)$ and $C(z) =C_0 + \delta C(z)$, where $L_0$ and $C_0$ are unperturbed constant inductance and capacitance per unit length, and $|\delta C| \ll C_0, |\delta L| \ll L_0$ are defined as the small perturbations in capacitance and inductance per unit length of the transmission line as functions of position $z$ around the ring. Eq.\ref{eqn:S2_telegrapher_equation} can then be cast into the form
\begin{equation} \label{eqn:S2_reduced_telegrapher_equation_with_perturbation}
\begin{split}
&\frac{\partial V}{\partial t} \approx  - \frac{1}{C_0} \left( 1 - \frac{\delta C(z)}{C_0} \right) \frac{\partial I}{\partial z}, \\
&\frac{\partial I}{\partial t} \approx  - \frac{1}{L_0} \left( 1 - \frac{\delta L(z)}{L_0} \right) \frac{\partial V}{\partial z}.
\end{split}
\end{equation}
Eq.\ref{eqn:S2_reduced_telegrapher_equation_with_perturbation} is the `Hamiltonian' form of transmission line equations \cite{louisell1960coupled}. For small inhomogeneities and high-Q resonators, the lowest-order modes of interest are expected to be similar to that of a uniform ring. Higher-order harmonic modes also couple neighbouring ports, e.g., at twice or four times the lowest-order resonance frequency, but their cross-mode contributions into the lowest-order modes are minimal for high-Q ring resonators. 

The general solution can be expanded in terms of the lowest-order modes of the perturbed ring using Eq.\ref{eqn:S2_general_solution_of_telegrapherequation_decompostion_1st_order}. Substituting Eq.\ref{eqn:S2_general_solution_of_telegrapherequation_decompostion_1st_order} into Eq.\ref{eqn:S2_reduced_telegrapher_equation_with_perturbation} and utilizing the orthonormality of $\sin \left( z/r \right)$ and $\cos \left( z/r \right)$ with respect to their inner product, Eq.\ref{eqn:S2_reduced_telegrapher_equation_with_perturbation} can be decomposed into the following equations describing the time evolution of decomposition coefficients
\begin{equation}\label{eqn:S2_perturbed_modes_form}
\begin{split}
&\frac{\partial \mathbf{u}_a}{\partial t} = i \omega_0 \Omega_1 \mathbf{u}_a + i \omega_0 \Omega_2 \mathbf{u}_b, \\
& \frac{\partial \mathbf{u}_b}{\partial t} = -i \omega_0 \Omega_1 \mathbf{u}_b - i \omega_0 \Omega_2 \mathbf{u}_a,
\end{split}
\end{equation}
where $\omega_0 = 1/(r\sqrt{L_0 C_0})$ is identified as the resonance frequency of the unloaded ring resonator and 
%
\begin{equation}\label{eqn:S2_perturbed_matrix}
\begin{split}
  \mathbf{u}_a &= \frac{1}{\sqrt{2}} \left[ \mathbf{u}_l + i \begin{bmatrix} 0 & 1\\ -1 & 0 \end{bmatrix} \mathbf{u}_c \right], \\
      \mathbf{u}_b &= \frac{1}{\sqrt{2}} \left[ \mathbf{u}_l - i \begin{bmatrix} 0 & 1\\ -1 & 0 \end{bmatrix} \mathbf{u}_c \right],  \\ 
    \Omega_1 &= \begin{bmatrix} 1 - (l_{cc} + c_{ss})/2 & (c_{cs} - l_{cs})/2\\ (c_{cs} - l_{cs})/2 & 1 - (c_{cc} + l_{ss})/2 \end{bmatrix}\,,\\ 
    \Omega_2 &= \begin{bmatrix} ( l_{cc} - c_{ss} )\slash 2& ( c_{cs} + l_{cs} )\slash 2 \\ ( c_{cs} + l_{cs} )\slash 2& ( l_{ss} + c_{cc} )\slash 2 \end{bmatrix} \, ,\\
    \mathbf{u}_l &= \sqrt{\frac{\pi r L_0}{2}} \begin{bmatrix} I_c (t)\\ I_s (t) \end{bmatrix},   \\
    \mathbf{u}_c &= \sqrt{\frac{\pi r C_0}{2}} \begin{bmatrix} V_c (t)\\ V_s (t) \end{bmatrix}.
\end{split}
\end{equation}
%
%
Here $|\mathbf{u}_l|^2$ and $|\mathbf{u}_c|^2$ give the total energy stored in the capacitive and inductive parts of the line, respectively,  and satisfy $|\mathbf{u}_a|^2 + |\mathbf{u}_b|^2 = |\mathbf{u}_l|^2 + |\mathbf{u}_c|^2$. $\mathbf{u}_a$ and $\mathbf{u}_b$ correspond to a set of lightly perturbed modes with $\text{exp}(i \omega_0 t)$ positive frequency dependence and $\text{exp}(- i \omega_0 t)$ negative frequency dependence, respectively.  The components of each vector indicate the relative amplitudes of two spatially orthogonal modes.  The quantities of the form $l_{ss}$, $l_{cs}$ are given by
\begin{equation}
\begin{split}
& l_{ss} = \frac{1}{\pi r} \int_0^{2 \pi r} \frac{\delta L(z)}{L_0} \sin^2 \left( \frac{z}{r} \right) dz,\\
& l_{cc} = \frac{1}{\pi r} \int_0^{2 \pi r} \frac{\delta L(z)}{L_0} \cos^2 \left( \frac{z}{r} \right) dz, \\
&c_{ss} = \frac{1}{\pi r} \int_0^{2 \pi r} \frac{\delta C(z)}{C_0} \sin^2 \left( \frac{z}{r} \right) dz,\\
&c_{cc} = \frac{1}{\pi r} \int_0^{2 \pi r} \frac{\delta C(z)}{C_0} \cos^2 \left( \frac{z}{r} \right) dz,\\
&l_{cs} = \frac{1}{\pi r} \int_0^{2 \pi r} \frac{\delta L(z)}{L_0} \cos \left( \frac{z}{r} \right) \sin \left( \frac{z}{r} \right) dz,\\
&c_{cs} = \frac{1}{\pi r} \int_0^{2 \pi r} \frac{\delta C(z)}{C_0} \cos \left( \frac{z}{r} \right) \sin \left( \frac{z}{r} \right) dz,
\end{split} 
\end{equation}
and correspond to the amplitudes of different spatial components in the inhomogeneities.

In the absence of perturbations,  following the same mode construction method as in the above perturbed case,  Eq.\ref{eqn:S2_telegrapher_equation} reduces to
%
\begin{align} \label{eqn:S2_mode_form_ideal_lossless_ring}
\frac{\partial \mathbf{u}}{\partial t} &= i \omega_0 \Omega_0 \cdot \mathbf{u},  \\
\mathbf{u} = \frac{\sqrt{\pi}}{2} \resizebox{0.9\width}{!}{$  \begin{bmatrix} i& 1& 0& 0\\ -i& 1& 0& 0\\ 0& 0& 1& -i\\ 0& 0& 1& i\end{bmatrix}$} & \resizebox{0.9\width}{!}{$\begin{bmatrix} \sqrt{rL}\, I_s\\ \sqrt{rC}\, V_c\\ \sqrt{rC}\, V_s\\ \sqrt{rL}\, I_c \end{bmatrix}$}, \,\, \Omega_0 = \resizebox{0.9\width}{!}{$\begin{bmatrix} 1& 0& 0& 0\\  0& -1& 0& 0\\ 0& 0& 1& 0\\ 0& 0& 0& -1\\ \end{bmatrix}$}.  \notag
\end{align}

Eq.\ref{eqn:S2_mode_form_ideal_lossless_ring} is the modal form $\partial_t \mathbf{u} = i \Omega \cdot \mathbf{u}$ commonly used in temporal coupled-mode theory (TCMT) \cite{haus1984waves, zhao2019connection, fan2003temporal, suh2004temporal},  and emphasises that a set of orthogonal modes can be excited independently in the ring resonator.  

\subsection{MODE SPLITTING AND ROTATION }\label{S2_subsec_mode_splitting_and_rotation}
Real devices will always have small defects due to imperfections during the fabrication process, or non-uniformities in the metal or dielectrics which may break the symmetry and result in frequency splitting between degenerate modes \cite{makimoto2013microwave, gopalakrishnan1990bandpass}. When such inhomogeneities are present, as represented by $\delta L$ and $\delta C$ in our model, the modes in Eq.\ref{eqn:S2_mode_form_ideal_lossless_ring} are no longer the eigenmodes and a new basis set needs to be calculated using Eq.\ref{eqn:S2_perturbed_modes_form}. In high-Q resonators, the coupling between positive and negative frequency modes is weak and can be ignored to first order under the coupled-mode approximation \cite{louisell1960coupled}. We neglect $\mathbf{u}_b$ terms in the equations for $\mathbf{u}_a$ and vice-versa, Eq.\ref{eqn:S2_perturbed_modes_form} can be simplified as
\begin{equation}\label{eqn:S2_perturbed_modes_form_without_+-_coupling}
\frac{\partial \mathbf{u}_a}{\partial t} = i \omega_0 \Omega_1 \mathbf{u}_a,\,\,\,\,\,\frac{\partial \mathbf{u}_b}{\partial t} = -i \omega_0 \Omega_1 \mathbf{u}_b.
\end{equation}

For small perturbations of inductance and capacitance along the line, the characteristic impedance $Z = \sqrt{L/C}$ and wave speed $v_p = 1/\sqrt{LC}$ approximately vary as 
\begin{equation}
\begin{split}
    \delta Z/Z_0 &\approx \delta L/2 L_0  - \delta C/2 C_0 \,, \,\,\,\,  Z_0 = \sqrt{L_0/C_0}\,,\\
    \delta v_p/v_{p_0} &\approx - \delta L/2 L_0  - \delta C/2 C_0\,, \,\,\, v_{p_0} = 1/\sqrt{L_0 C_0}\,. 
\end{split}
\end{equation}
%
For low loss, the solutions obtained by explicit diagonalisation and by the coupled-mode approximation are identical \cite{louisell1960coupled}.  The resonance frequencies $\omega_{a_1}$ and $\omega_{a_2}$ of the normal modes of the perturbed system are given by the eigenvalues of the matrix $\omega_0 \Omega_1$.  The calculation yields $\omega_{a_1} = (1 + \alpha) \omega_0 + \Delta \omega/2$ and $\omega_{a_2} = (1 + \alpha) \omega_0 - \Delta \omega/2$, where resonance frequency shift $\alpha$ and frequency splitting integral $\Delta \omega/\omega_0$ are given by
\begin{equation}\label{eqn:S2_frequency_shift_and_splitting}
\begin{split}
\alpha &= \frac{1}{2 \pi r} \int_0^{2 \pi r} \frac{\delta v_p}{v_{p_0}} \, dz,\\ 
\frac{\Delta \omega}{\omega_0} &= \frac{1}{\pi r} \int_0^{2 \pi r} \frac{\delta Z}{Z_0} \text{exp} \left( \frac{2 i z}{r} \right) dz.
\end{split}
\end{equation}
The magnitude of fractional frequency splitting is quantified by the modulus of the above integral. Eq.\ref{eqn:S2_frequency_shift_and_splitting} shows that the perturbations in wave speed around the ring shift the resonance frequency compared to a uniform ring, but do not cause frequency splitting between the two lowest-order modes. Perturbations in characteristic impedance do cause frequency splitting between modes, and the fractional frequency splitting is proportional to the root-mean-square fluctuation in the line characteristic impedance around the ring. The form of the integral indicates that random impedance fluctuations on length scales much smaller than a ring quadrant have a small effect, as their contributions tend to average out. However, systematic shifts in impedance introduced during fabrication can produce impedance gradients that lead to significant frequency splitting. Similarly, localised defects that result in a large ${\delta Z}/{Z_0}$ in a small region of space can also lead to significant frequency splitting.

The re-diagonalisation of the matrices in Eq.\ref{eqn:S2_perturbed_matrix} is achieved by rotating the modes relative to the ports by an angle $\theta_0$, given by
\begin{equation}\label{eqn:S2_mode_rotation_angle_theta-0}
\theta_0 = \mathrm{arctan} \left( \frac{- \Im [ \Delta \omega]}{\Re[ \Delta \omega]  \pm |\Delta \omega| }  \right),
\end{equation}
where $\Im[\Delta \omega]$ and $\Re[\Delta \omega]$ denote taking the imaginary and the real components of $\Delta \omega$ respectively. Eq.\ref{eqn:S2_mode_rotation_angle_theta-0} shows that the perturbation in characteristic impedance produces both frequency splitting and a rotation of the modes relative to the ports. This rotation significantly reduces the isolation between the two channels.  

In summary, artifacts, defects, or other non-uniformities such as external fields give rise to localised or distributed variations in $v_p$ and $Z$. As described by equations (\ref{eqn:S2_frequency_shift_and_splitting}) and (\ref{eqn:S2_mode_rotation_angle_theta-0}), these variations manifest at the device level as frequency splitting and mode rotation.

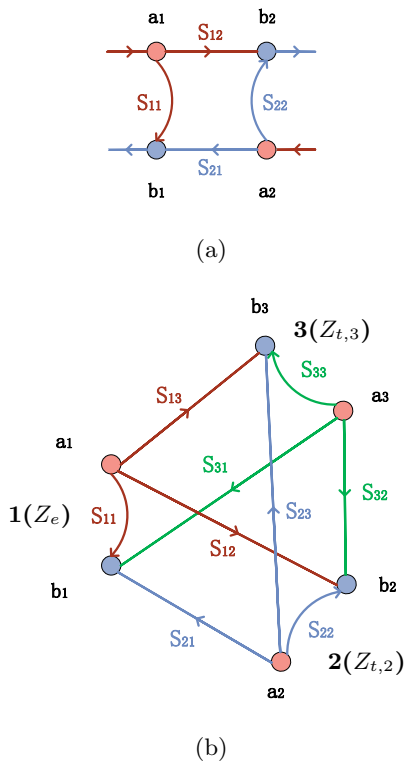
\begin{figure}[H]
\centering
\begin{subfigure}[b]{0.50\textwidth}
\centering
    \begin{tikzpicture}
        \node[inner sep=0, xshift=0cm, yshift=0cm] (image) at (0,0) {\includegraphics[width = .4\linewidth]{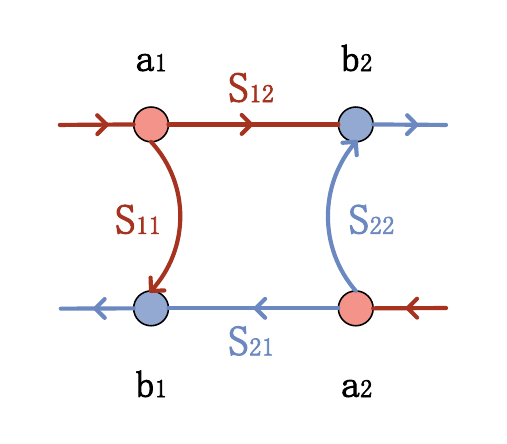}};
     \end{tikzpicture}
   \caption{}
\end{subfigure}

\begin{subfigure}[b]{0.50\textwidth}
\centering
    \begin{tikzpicture}
        \node[inner sep=0, xshift=0cm, yshift=0.25cm] (image) at (0,0) {\includegraphics[width= .6\linewidth]{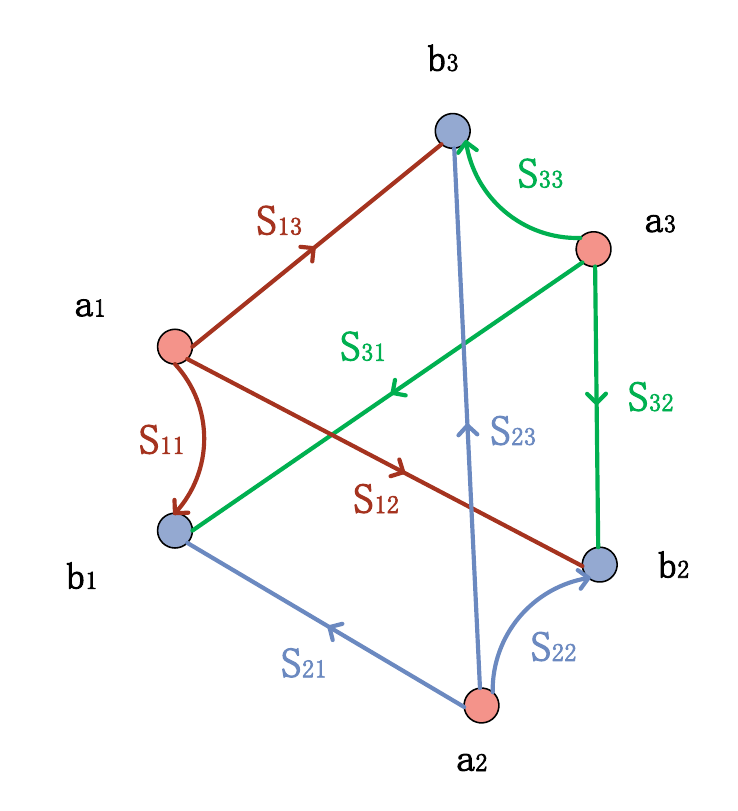}};
        \node[text=black, scale=1, align=left, font=\bfseries] at (-2.35, 0) {1($Z_e$)};
        \node[text=black, scale=1, align=left, font=\bfseries] at (1.95, -1.95) {2($Z_{t,2}$)};
        \node[text=black, scale=1, align=left, font=\bfseries] at (1.5, 2.45) {3($Z_{t,3}$)};
     \end{tikzpicture}
   \caption{}
   
\end{subfigure}
\caption{Flow graph representation of (a) a two-port waveguide, and (b) a three-port coupler.}
\label{fig:S2_flow_graph_components}
\end{figure}

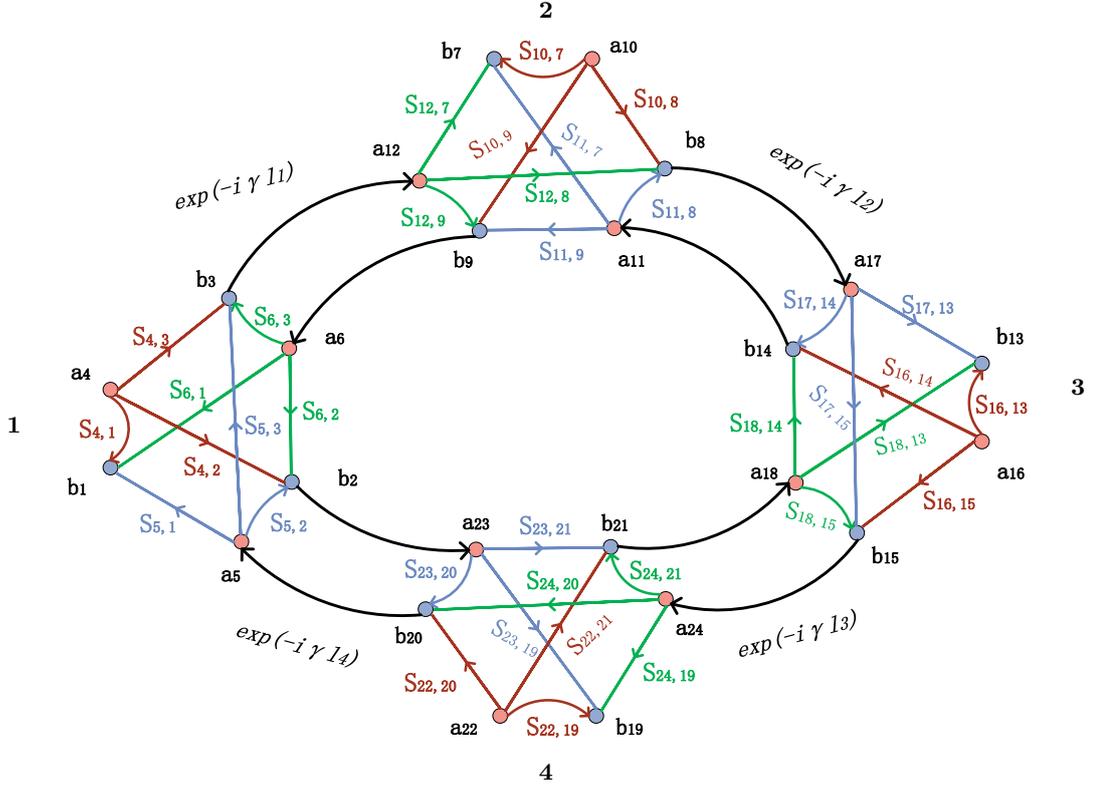
\begin{figure*}
\centering
\vspace{-0.5cm}
    \begin{tikzpicture}
        \node[inner sep=0, xshift=0cm, yshift= 0cm] (image) at (0,0) {\includegraphics[width=.75\linewidth]{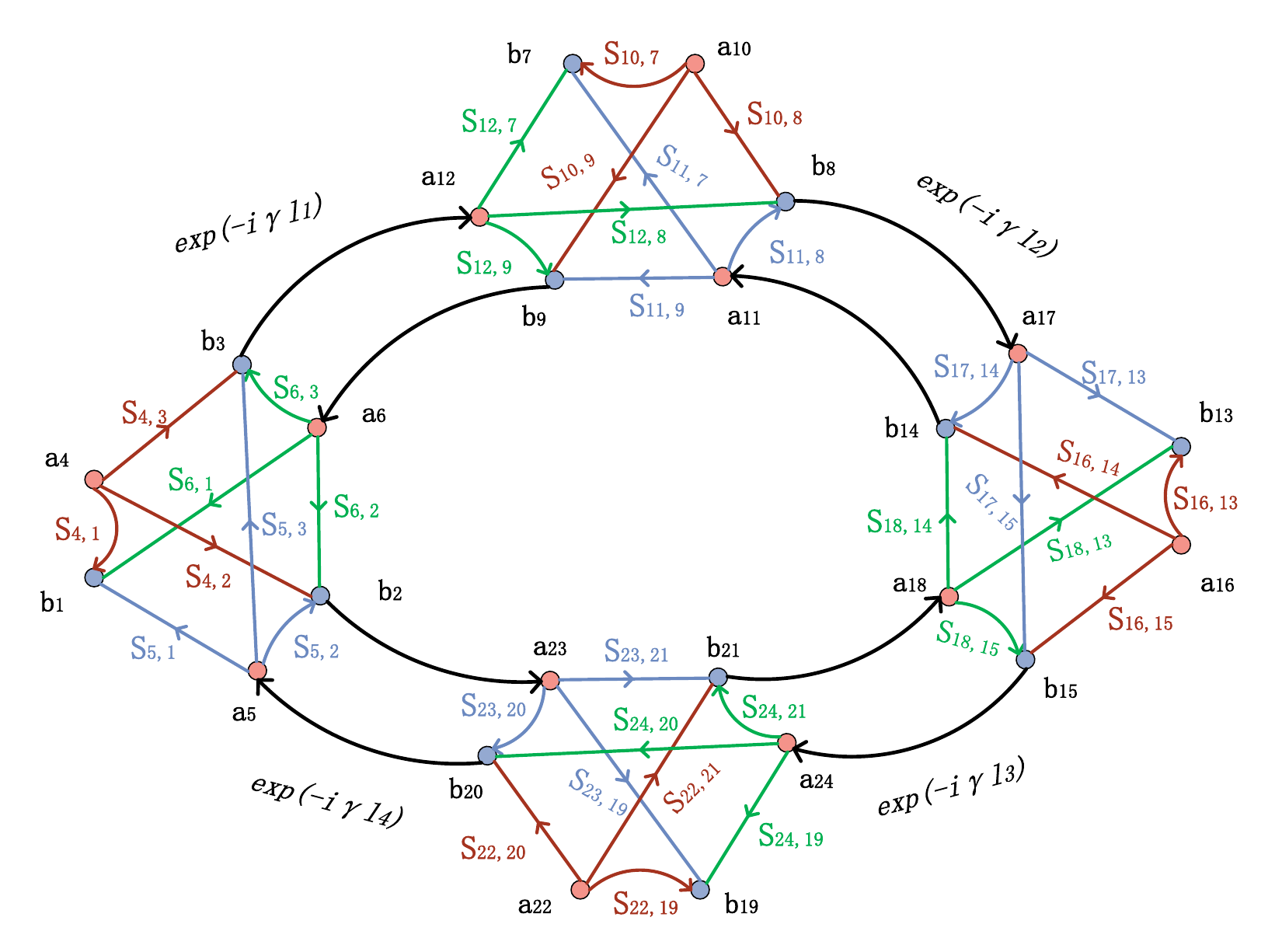}};
        
        \node[text=black, scale=1, align=left, font=\bfseries] at (-7, -.5) {1};
        \node[text=black, scale=1, align=left, font=\bfseries] at (0, 5) {2};
        \node[text=black, scale=1, align=left, font=\bfseries] at (7, 0) {3};
        \node[text=black, scale=1, align=left, font=\bfseries] at (0, -5.1) {4};
     \end{tikzpicture}
 \caption[ ]{\label{fig:S2_flow_graph_four-port_ring}  Flow graph of four-port ring resonator.}
\end{figure*}
\subsection{FLOW GRAPH SIMULATIONS}\label{sec_S3_simulation}
To verify our analytic results, we carried out numerical simulations using the flow graph method previously developed in \cite{withington2013elastic, Djelal2016thermal}. This is a robust analysis method which enables us to study the effect of asymmetry in the coupling ports of the ring resonator system, as well as inhomogeneities along the transmission line. 

In our implementation of the flow graph simulations, the superconducting ring was divided into connected scattering sub-networks, each modelled in terms of its scattering parameters. These sub-networks were then linked by a connection matrix that captures the steady-state behaviour of the overall ring system.

Fig.\ref{fig:S2_flow_graph_components}(a) shows a two-port waveguide, each port of which comprises two nodes connected by directed branch and representing complex signal amplitudes travelling in opposite directions.  $a_i$ for incoming and $b_i$ for outgoing waves at the $i^{th}$ port, respectively. $S_{ij} = b_j / a_i$ is defined as the transmission coefficient from port $i$ to $j$ and $S_{ii} = b_i/a_i$ is the reflection coefficient at port $i$ itself.  

The scattering matrix of the two-port waveguide is given by 
\begin{equation}
\begin{bmatrix}
b_1\\ b_2
\end{bmatrix} = \begin{bmatrix}
S_{11} & S_{21}\\ S_{12} & S_{22}
\end{bmatrix} \begin{bmatrix}
a_1\\ a_2
\end{bmatrix},
\end{equation}
which satisfies two constraints: reciprocity $S_{ij} = S_{ji}$ and symmetry $S_{ij} = s$ and $S_{ii} = r$. Consequently, there are only two independent values $s$ and $r$ in the matrix. 

\begin{figure*}
\centering
\begin{tikzpicture}
      \node[inner sep=0, xshift=0cm, yshift=0cm] (image) at (0,0) {\includegraphics[width=\linewidth]{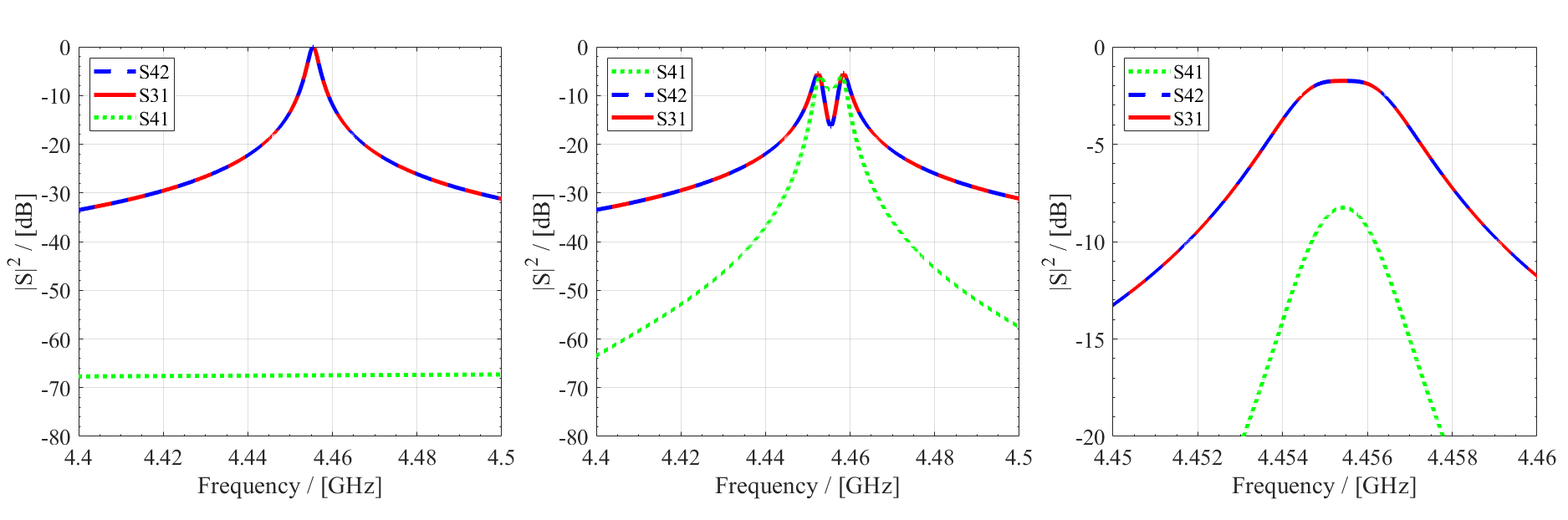}};
            \node[text=black, scale=1, align=left] at (-6,-3) {(a)};
      \node[text=black, scale=1, align=left] at (0,-3) {(b)};
      \node[text=black, scale=1, align=left] at (6,-3) {(c)};
      
\end{tikzpicture}
\caption{\label{fig:c1_compiled_splitting} Flow graph simulations of ring resonators with identical coupling capacitances at all external ports $C_c = 20.8\,fF$. S42 and S31 correspond to the transmission between two orthogonal pairs of opposite ports, while S41 corresponds to the transmission between adjacent ports. (a) No perturbation on the ring transmission lines.  (b) Quadrupole impedance perturbation on the ring transmission lines: lines 1 and 3 have $Z_{t}=50.05\,\Omega$ and lines 2 and 4 have $Z_{t}=49.95\,\Omega$.  (c) Weaker quadrupole impedance perturbation on the ring transmission lines: lines 1 and 3 have $Z_{t}=50.01\,\Omega$ and lines 2 and 4 have $Z_{t}=49.99\,\Omega$.}
\end{figure*}

Each port of the ring resonator includes a coupling capacitor and can be modelled as a lossless reciprocal three-port coupler, as shown in Fig.\ref{fig:S2_flow_graph_components}(b). Here, port 1 connects to the external transmission line with characteristic impedance $Z_e$ via a coupling capacitor connected in series with impedance $Z_c = 1/i \omega C_c$.  Port 2 and port 3 connect to the ring line with characteristic impedances $Z_{t,2}$ and $Z_{t,3}$ respectively. The scattering matrix of a ring port is obtained by extending the symmetric case in Ref.\cite{guruswamy2018nonequilibrium}, where $Z_{t,2}=Z_{t,3}$, to the more general case of $Z_{t,2}\neq Z_{t,3}$, and it is given by
\begin{equation}\label{eqn:S3_three_port_coupler_matrix}
S_p =  \begin{bmatrix}
S11 & S12 & S13\\
S21 & S22 & S23\\
S31 & S32 & S33
\end{bmatrix}\,,
\end{equation}
where 
\begin{align}
    S11 &= \frac{Z_c+Z_{2,3}-Z_e}{Z_c+Z_{2,3}+Z_e}\,,\notag \\
    S22 &= \frac{Z_{1,3}-Z_{t,2}}{Z_{1,3}+Z_{t,2}}\,,\notag \\
    S33 &= \frac{Z_{1,2}-Z_{t,3}}{Z_{1,2}+Z_{t,3}}\,,\notag \\
    S21 &= S12 = 2\sqrt{\frac{Z_e}{Z_{t,2}}}\frac{Z_{2,3}}{Z_{e}+Z_{c}+Z_{2,3}}\,,\notag \\
    S31 &= S13 = 2\sqrt{\frac{Z_e}{Z_{t,3}}}\frac{Z_{2,3}}{Z_{e}+Z_{c}+Z_{2,3}}\,,\notag \\
    S32 &= S23 = 2\sqrt{\frac{Z_{t,3}}{Z_{t,2}}}\frac{Z_{1,2}}{Z_{t,3}+Z_{1,2}}\,,\notag
\end{align}
and the parallel impedances are given by
\begin{align}
    (Z_{2,3})^{-1} &= Z_{t,2}^{-1}+Z_{t,3}^{-1}\,,\notag \\
    (Z_{1,2})^{-1} &= Z_{t,2}^{-1}+(Z_c+Z_e)^{-1}\,,\notag \\
    (Z_{1,3})^{-1} &= Z_{t,3}^{-1}+(Z_c+Z_e)^{-1}\,.\notag
\end{align}

Building on the two basic components above, we construct the overall connection matrix $P$ for the ring resonator, which incorporates all scattering coefficients between adjacent nodes within each basic element, as well as the wave propagation between elements, as shown in Fig.\ref{fig:S2_flow_graph_four-port_ring}. We first collect all nodes in the network into a single vector
\begin{equation}
\begin{split}
    \mathbf{v} &= [ B_1\,  A_1 \, B_2 \,  A_2 \,   B_3 \,  A_3 \,  B_4 \,  A_4 ]^T\,,\\
    A_1 &= [a_4 \,  a_5 \,  a_6]^T\,, \,\,\, \,\,\,\,\,\,\,\,B_1 = [b_1 \,  b_2 \,  b_3]^T \,,  \\
    A_2 &= [a_{10} \,  a_{11} \,  a_{12}]^T \,, \,\,\, B_2 = [b_7 \,  b_8 \,  b_9]^T \,,  \\
   A_3 &= [a_{16} \,  a_{17} \,  a_{18}]^T \,,  \,\,\, B_3 = [b_{13} \,  b_{14} \,  b_{15}]^T \,,  \\
   A_4 &= [a_{22} \,  a_{23} \,  a_{24}]^T\,,\,\,\,  B_4 = [b_{19} \,  b_{20} \,  b_{21}]^T \,.
\end{split} 
\end{equation}

The waveguide connecting external port 1 and port 2 of the ring resonator has length $l_1$ and propagation constant $\gamma$, which can be determined using $\gamma=i \omega \sqrt{L C}$ as described in the transmission line model in Sec.\ref{subsec_S2_model_of_ring_resonator}. The corresponding mapping relations is given by $A_2 = S_w^3(1) B_1$.  Here, $S_w^{j}(i)$ indicates the scattering matrix of the $i_{th}$ ring quadrant waveguide with an entry only at the superscript-$j$, given by $S_w^3(n) = \text{diag}(0, 0, e^{-i \gamma l_n})$ and $S_w^2(n) = \text{diag}(0, e^{-i \gamma l_n}, 0)$. The output waves of the three-port coupler at external port 1 can be calculated using $B_1 = S_p(C_{c1}) A_1$, where $C_{c1}$ is the coupling capacitance at external port 1. By repeating the same procedure at all other ports, the full connection matrix $P$ is obtained
\begin{equation*}
\resizebox{\hsize}{!}{$P=\begin{bmatrix}
0& S_p(C_{c1}) &0 &0 &0 &0 &0 &0\\
0& 0&S_w^3(1) & 0& 0& 0& S_w^2(4)&0\\
0&0 &0 &S_p(C_{c2}) &0 &0 &0 &0\\
S_w^3(1) &0 &0 &0 &S_w^2(2) &0 & 0&0\\
0&0 &0 &0 &0 &S_p(C_{c3}) &0 &0\\
0& 0&S_w^2(2) &0 &0 &0 &S_w^3(3) &0\\
0&0 & 0& 0&0 & 0& 0&S_p(C_{c4})\\
S_w^2(4)&0 &0 & 0&S_w^3(3) &0 &0 &0\\
\end{bmatrix},$}
\end{equation*}
which must satisfy $\mathbf{v} = P \mathbf{v}$.  The response of the ring resonator to external sources can be modelled by introducing a source amplitude vector $\mathbf{n}$ such that $(I - P) \cdot \mathbf{v} = \mathbf{n}$ \cite{withington2013elastic} . We define the matrix $Q = I - P$ to relate the vector of wave amplitudes against the vector of source amplitudes. Since $Q$ may be singular, we calculate $Q^+$, the Moore-Penrose pseudo-inverse of $Q$, which is unique for a singular matrix and the inverse for a non-singular matrix. We then get $\mathbf{v} = Q^+ \mathbf{n}$, which projects the vector of sources to the vector of wave amplitudes at every node. Ignoring the $Q^+$ elements associated with internal nodes, we obtain the scattering matrix for the four-port ring
\begin{equation}\label{eqn:S2_S_prameters_flowgraph_matrix}
S = \begin{bmatrix}
Q^+_{1,4} & Q^+_{1,10}  & Q^+_{1,16}  & Q^+_{1,22} \\
Q^+_{7,4} & Q^+_{7,10}  & Q^+_{7,16}  & Q^+_{7,22} \\
Q^+_{13,4} & Q^+_{13,10}  & Q^+_{13,16}  & Q^+_{13,22} \\
Q^+_{19,4} & Q^+_{19,10}  & Q^+_{19,16}  & Q^+_{19,22} 
\end{bmatrix}.
\end{equation}

\begin{figure*}[htbp!]
\centering
\begin{tikzpicture}
  	   \node[inner sep=0, xshift=0cm, yshift=0cm] (image) at (0,0) {\includegraphics[width=\linewidth]{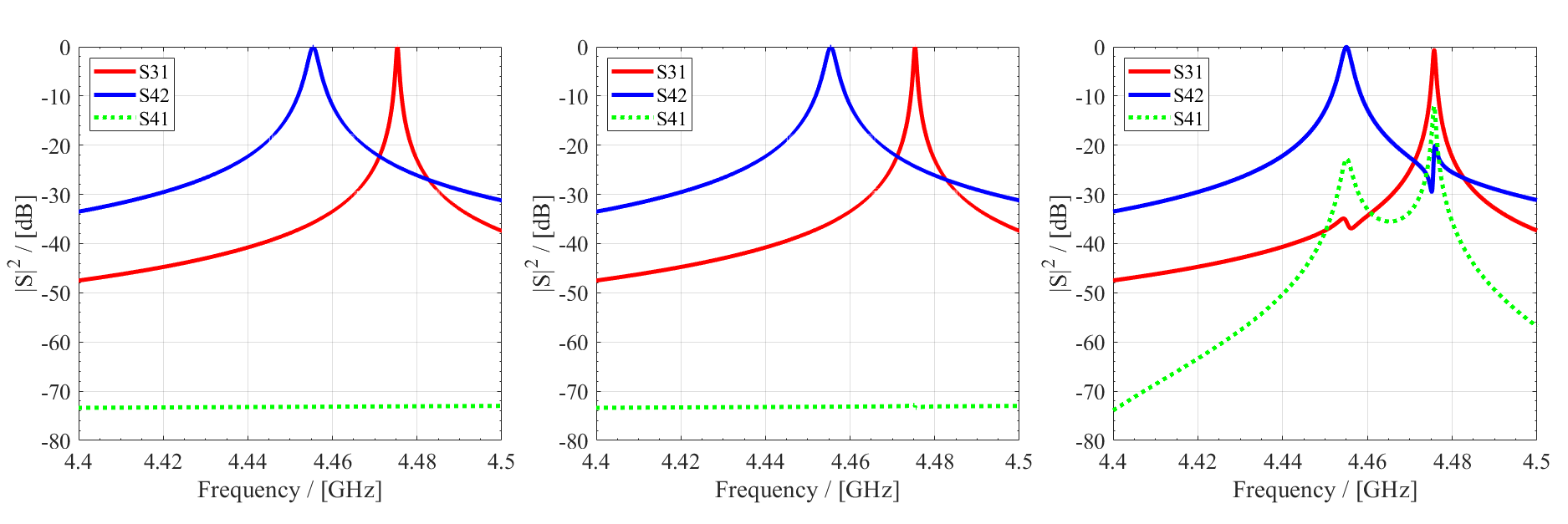}};
  	   \node[text=black, scale=1, align=left] at (-6,-3) {(a)};
    	\node[text=black, scale=1, align=left] at (0,-3) {(b)};
    	\node[text=black, scale=1, align=left] at (6,-3) {(c)};
  	   
  	   \begin{scope}[shift={(3,4)}]
    	 \draw[fill=white,line width = 0.025cm] (-1.5,-4.75) circle (0.6);  
    	 \draw[thin] (-2.1, -4.75) -- (-0.9, -4.75);
         \draw[thin] (-1.5, -5.35) -- (-1.5, -4.15);
        
           	 \draw[fill=white, line width = 0.02cm] (-2.3,-4.75) circle (0.15);  
           	 \node[text=black, scale=1, align=left] at (-2.3,-4.75) {1};
           	 \draw[fill=white, line width = 0.02cm] (-0.7,-4.75) circle (0.15);  
           	\node[text=black, scale=1, align=left] at (-0.7,-4.75) {3};
           	\draw[fill=white, line width = 0.02cm] (-1.5,-3.95) circle (0.15);  
           	\node[text=black, scale=1, align=left] at (-1.5,-3.95) {2};
           	\draw[fill=white, line width = 0.02cm] (-1.5,-5.55) circle (0.15);  
           	\node[text=black, scale=1, align=left] at (-1.5,-5.55) {4};
           	
           	 \node[text=black, scale=.75, align=left] at (-2,-4.2) {\textit{1}};
           	\node[text=black, scale=.75, align=left] at (-1.05,-5.3) {\textit{3}};
           	\node[text=black, scale=.75, align=left] at (-1.05,-4.2) {\textit{2}};
           	\node[text=black, scale=.75, align=left] at (-2,-5.3) {\textit{4}};
           	
         \draw[line width = 0.05cm] (-1.7, -4.45) -- (-1.9, -4.45);
         \draw[line width = 0.05cm] (-1.1, -4.45) -- (-1.3, -4.45);
         \draw[line width = 0.05cm] (-1.7, -5.05) -- (-1.9, -5.05);
         \draw[line width = 0.05cm] (-1.2, -4.95) -- (-1.2, -5.15);
         
         \draw[line width = 0.05cm] (-1.1, -5.05) -- (-1.3, -5.05);
         \draw[line width = 0.05cm] (-1.8, -4.95) -- (-1.8, -5.15);
         \end{scope}
         \begin{scope}[shift={(8.5,4)}]
              \draw[fill=white, line width = 0.02cm] (-2.3,-4.75) circle (0.15);  
           	 \node[text=black, scale=1, align=left] at (-2.3,-4.75) {1};
           	 \draw[fill=white, line width = 0.02cm] (-0.7,-4.75) circle (0.15);  
           	\node[text=black, scale=1, align=left] at (-0.7,-4.75) {3};
           	\draw[fill=white, line width = 0.02cm] (-1.5,-3.95) circle (0.15);  
           	\node[text=black, scale=1, align=left] at (-1.5,-3.95) {2};
           	\draw[fill=white, line width = 0.02cm] (-1.5,-5.55) circle (0.15);  
           	\node[text=black, scale=1, align=left] at (-1.5,-5.55) {4};
           	
           	 \node[text=black, scale=.75, align=left] at (-2,-4.2) {\textit{1}};
           	\node[text=black, scale=.75, align=left] at (-1.05,-5.3) {\textit{3}};
           	\node[text=black, scale=.75, align=left] at (-1.05,-4.2) {\textit{2}};
           	\node[text=black, scale=.75, align=left] at (-2,-5.3) {\textit{4}};
         \end{scope}
         \begin{scope}[xscale = -1, shift={(-13.5, 4)}, ]
    	\draw[fill=white, line width = 0.025cm] (6.5,-4.75) circle (0.6);  
    	\draw[thin] (5.9, -4.75) -- (7.1, -4.75);
        \draw[thin] (6.5, -5.35) -- (6.5, -4.15);

         \draw[line width = 0.05cm] (6.7, -4.45) -- (6.9, -4.45);
         \draw[line width = 0.05cm] (6.8, -4.35) -- (6.8, -4.55);
         \draw[line width = 0.05cm] (6.1, -4.45) -- (6.3, -4.45);
         \draw[line width = 0.05cm] (6.7, -5.05) -- (6.9, -5.05);
         \draw[line width = 0.05cm] (6.1, -5.05) -- (6.3, -5.05);
      \draw[line width = 0.05cm] (6.2, -4.95) -- (6.2, -5.15);
      \end{scope}
\end{tikzpicture}
\caption{\label{fig:c2_compiled_Z_perturbation} Flow graph simulations of ring resonators with coupling capacitances of $C_c = 20.8\,fF$ at port 2 and 4, and coupling capacitances of $C_c = 10.8\,fF$ at port 1 and 3. S42 and S31 correspond to the transmission between two orthogonal pairs of opposite ports, while S41 corresponds to the transmission between adjacent ports. (a) No perturbation on the ring transmission lines.  (b) Dipole impedance perturbation on the ring transmission lines: lines 1 and 2 have $Z_{t}=49.95\,\Omega$ and lines 3 and 4 have $Z_{t}=50.05\,\Omega$.  (c) Quadrupole impedance perturbation on the ring transmission lines: lines 1 and 3 have $Z_{t}=50.05\,\Omega$ and lines 2 and 4 have $Z_{t}=49.95\,\Omega$.}
\end{figure*}

We have applied the flow graph method to a set of simulations shown in Fig.\ref{fig:c1_compiled_splitting} where a uniform, symmetric ring was capacitively loaded with four identical coupling ports, each featuring a capacitance of $C_c = 20.8\,fF$. Other parameters were set as follows: external transmission line impedance $Z_e = 50\,\Omega$, ring diameter $d =$ 10\,mm, and dielectric constant $\epsilon_r$ = 4.5. These parameter values were chosen to be close in size to microstrip ring devices that we fabricated and tested in our laboratory as shown in Fig.\ref{fig:S4_cpw_resonance_splitting_measurement}/\ref{fig:S4_microstrip_resonance_splitting_measurement}. We focus our discussions on S-parameters $S_{42}$, $S_{31}$ and $S_{41}$, which characterise transmissions across pairs of opposite ports, and into adjacent ports, respectively. Due to the uniformity in the capacitance of the ports in this first set of simulations, $S_{42}$ and $S_{31}$ are equal. We will explore cases where this uniformity is broken in a later set of simulations.

Fig.\ref{fig:c1_compiled_splitting}(a) shows the S-parameters in the absence of any impedance perturbation on the transmission lines, i.e. $Z_{t}=50\,\Omega$ for all transmission lines. $S_{42}$ and $S_{31}$ standard single-pole resonator resonance with resonance frequency at $f_r \sim$ 4.455\,GHz. There is a negligible transmission $S_{41}$ between adjacent ports which is more than $65\,\mathrm{dB}$ smaller compared to the peaks of $S_{42}$ and $S_{31}$. 

Fig.\ref{fig:c1_compiled_splitting}(b) shows the S-parameters when the impedances of the lines are perturbed, such that lines 1 and 3 have $Z_{t}=50.05\,\Omega$ and lines 2 and 4 have $Z_{t}=49.95\,\Omega$, i.e. in a quadrupole pattern. Although the amount of impedance perturbation is only one part in a thousand, the resonance has clearly split into distinct peaks several bandwidths apart. Further, $S_{41}$ now has peak transmission close to the peak of $S_{42}$ and $S_{31}$. This indicates that the isolation between adjacent ports has largely vanished in the presence of quadrupole impedance perturbations as small as one part in a thousand on the transmission lines. This highly sensitive frequency-splitting effect, together with the reduction in isolation, can be used experimentally to identify the presence of impedance perturbations. Importantly, it also highlights the need for careful control of line symmetry and impedance if the aim is to realise a near-ideal ring resonator. This high sensitivity to perturbations is partly attributed to the high Q-factor of the resonance, which in this case is $\approx2000$. A high Q-factor allows the resonance splitting to be resolved against the bandwidth of the resonance. Since superconducting resonators can exhibit Q-factors as high as $10^5$ or more, the effects of impedance perturbations can be particularly significant in superconducting ring resonators.

Fig.\ref{fig:c1_compiled_splitting}(c) shows the S-parameters when the impedance perturbation is reduced, such that lines 1 and 3 have $Z_{t}=50.01\,\Omega$ and lines 2 and 4 have $Z_{t}=49.99\,\Omega$. In this case, the perturbation does not result in complete splitting of the peaks. Instead a flattened transmission profile is obtained, which is the summation of two single-pole resonance profiles close in frequency. Such flattened transmission profiles could be useful in engineering broadened filter responses.

In this second set of simulations shown in Fig.\ref{fig:c2_compiled_Z_perturbation}, we have relaxed the uniformity in the capacitors such that port 2 and 4 have capacitances of $C_c = 20.8\,fF$ and port 1 and 3 have capacitances of $C_c = 10.8\,fF$. This breaks the symmetry in $S_{42}$ and $S_{31}$. As seen in Fig.\ref{fig:c2_compiled_Z_perturbation}(a), the resonance profiles of $S_{42}$ and $S_{31}$ now peak at different frequencies and have different bandwidths. Significantly, this asymmetry in resonance characteristics in the pairs of opposite ports does not result in mode splitting, mode rotation, or a reduction in isolation. As discussed in the previous section and shown in Eq.\ref{eqn:S2_frequency_shift_and_splitting}, the origin of these phenomena is in the integral of impedance perturbations along the transmission lines. In the absence of such perturbations, the modes remain pinned to the ports and no mixing occurs between them. 

In Fig.\ref{fig:c2_compiled_Z_perturbation}(b), we have introduced perturbation in the impedances of the transmission lines. In contrast to the previous case, here we have imposed that the perturbation has dipole pattern, such that lines 1 and 2 have $Z_{t}=49.95\,\Omega$ and lines 3 and 4 have $Z_{t}=50.05\,\Omega$. We notice that the S-parameters are nearly identical to that in Fig.\ref{fig:c2_compiled_Z_perturbation}(a), and no mode mixing occurs. This observation can be understood using Eq.\ref{eqn:S2_frequency_shift_and_splitting}, which shows that a perturbation with mirror symmetry leads to cancellation within the integral, resulting in no mode rotation. 

The lowest-order perturbation that yields a non-zero integral is the quadrupole pattern, which we have applied to obtain Fig.\ref{fig:c2_compiled_Z_perturbation}(c) by setting lines 1 and 3 to have $Z_{t}=50.05\,\Omega$ and lines 2 and 4 to have $Z_{t}=49.95\,\Omega$. As seen in the red and blue plots, $S_{42}$ and $S_{31}$ each exhibit a strong transmission peak at their respective resonance frequencies, accompanied by a weaker peak at the resonance frequency of the orthogonal channel. This indicates that the perturbed resonance modes are no longer pinned to the capacitive ports, but are instead rotated by a specific angle, as discussed previously in this paper. Further, the increase in the magnitude of the green plot indicates that there is a reduction in isolation between adjacent ports. This cross-talk now peaks at the resonance frequencies of the two orthogonal modes. In practice, to ensure frequency splitting remains insignificant compared to the bandwidth, condition $\delta \omega / \omega_0 \propto \delta Z_{rms} / Z \ll 1/Q$ must be satisfied, as dictated by Eq.\ref{eqn:S2_frequency_shift_and_splitting}. This condition, however, may impose stringent requirement on the tolerance control of the line impedance during fabrication, as superconducting resonators can exhibit very high Q-factors depending on their design (for example, $Q > 10^5$). As a result, superconducting ring resonators with high Q-factors are thus prone to significant frequency splitting and mode rotation.

\section{Experimental results}
\begin{figure}[htbp!]
\centering
  %
    \begin{tikzpicture}
        \node[inner sep=0, xshift=0cm, yshift=0cm] (image) at (0,0) {\includegraphics[width = 1\linewidth]{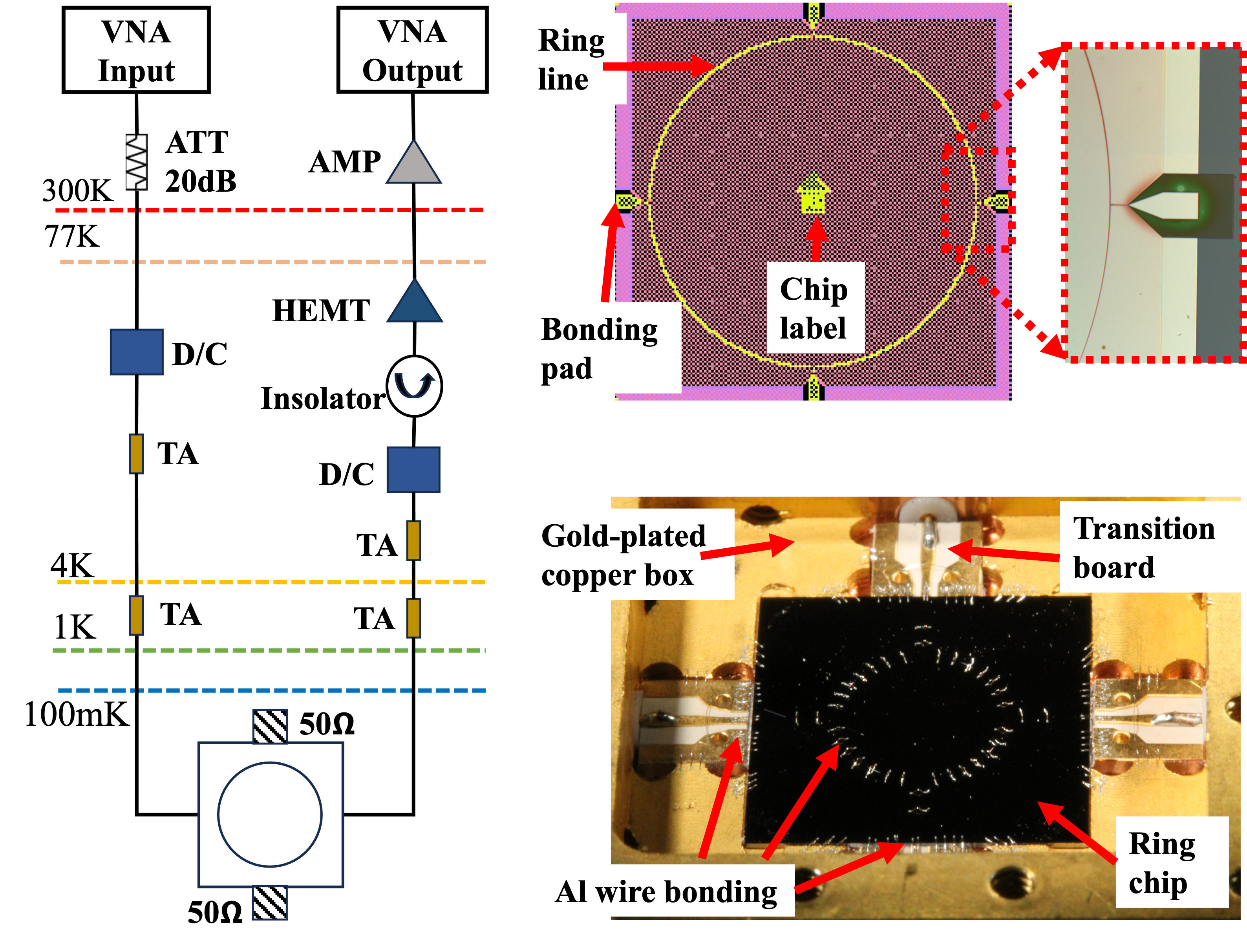}};
          \node[inner sep=0, xshift=0cm, yshift=0cm] at (-2.5,-3.45) {(a)};
           \node[inner sep=0, xshift=0cm, yshift=0cm] at (2.25,  0.25) {(b)};
            \node[inner sep=0, xshift=0cm, yshift=0cm] at (2.25,  -3.45) {(c)};
     \end{tikzpicture}
 \caption[ ]{\label{fig:S3_measurement_setup_and_ring_device} (a)Ring resonator measurement setup. VNA: vector network analyser, ATT: attenuator, D/C: DC block, TA: thermal anchor,  50$\Omega$: 50$\Omega$  matched load,  HEMT: high electron mobility transistor, AMP: amplifier at room temperature. (b)Ring chip mask. The zoomed-in section is a micrograph of the fabricated microstrip ring device. (c) Photograph of the packaged superconducting ring device.}
\end{figure}
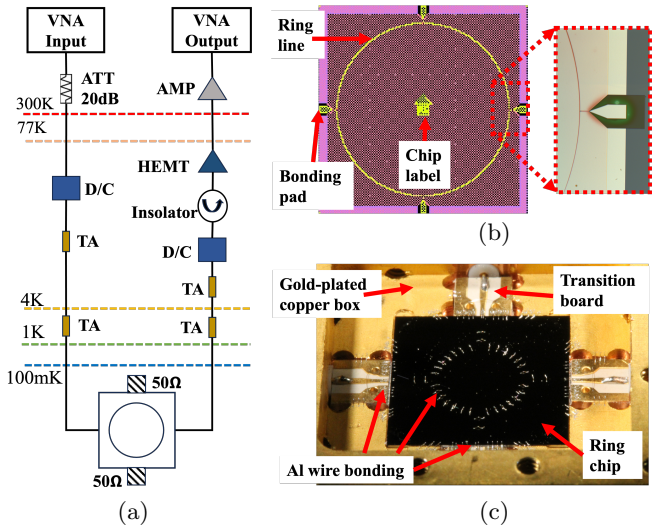

A set of Al and Nb ring resonators were designed and fabricated with microstrip and coplanar waveguide geometries, having T-junction parallel plates and open-ended series stub capacitors at the ports, respectively. The microstrip rings had a diameter of 10\,mm and line width of 2\,$\mu$m, as shown in Fig.\ref{fig:S3_measurement_setup_and_ring_device}(b). The microstrip line, deposited on 225\,$\mu$m thick Si wafer, consisted of 100\,nm thick top conductor, 500\,nm SiO$_2$ dielectric, and 150\,nm thick ground conductor. The CPW rings had a diameter of 6\,mm, the 100\,nm thick centre strip has a width of 5\,$\mu$m, and the gap width from conductor to ground was 6\,$\mu$m. The films were deposited using an ultra-high-vacuum magnetron sputtering system with a base pressure of $2 \times 10^{-10}$ Torr or lower. In the microstrip devices, Al layers were patterned by wet etching, Nb layers by reactive-ion etching, and the dielectric layers by a lift-off process. In the CPW devices, both Al and Nb layers were patterned by lift-off. The CPW ring centre plate was grounded by Al wires crossing the centre strip at each quadrant of the ring in a symmetrical pattern, as shown in Fig.\ref{fig:S3_measurement_setup_and_ring_device}(c). The coupling Q-factor was controlled by varying the overlap plate length and stub length over which gap coupling occurs, for microstrip and CPW rings, respectively. 

\begin{figure}[htbp!]
\centering
  	\begin{tikzpicture}
    	\node[inner sep=0, xshift=0cm, yshift=0cm] (image) at (0,0) {\includegraphics[width=.9\linewidth]{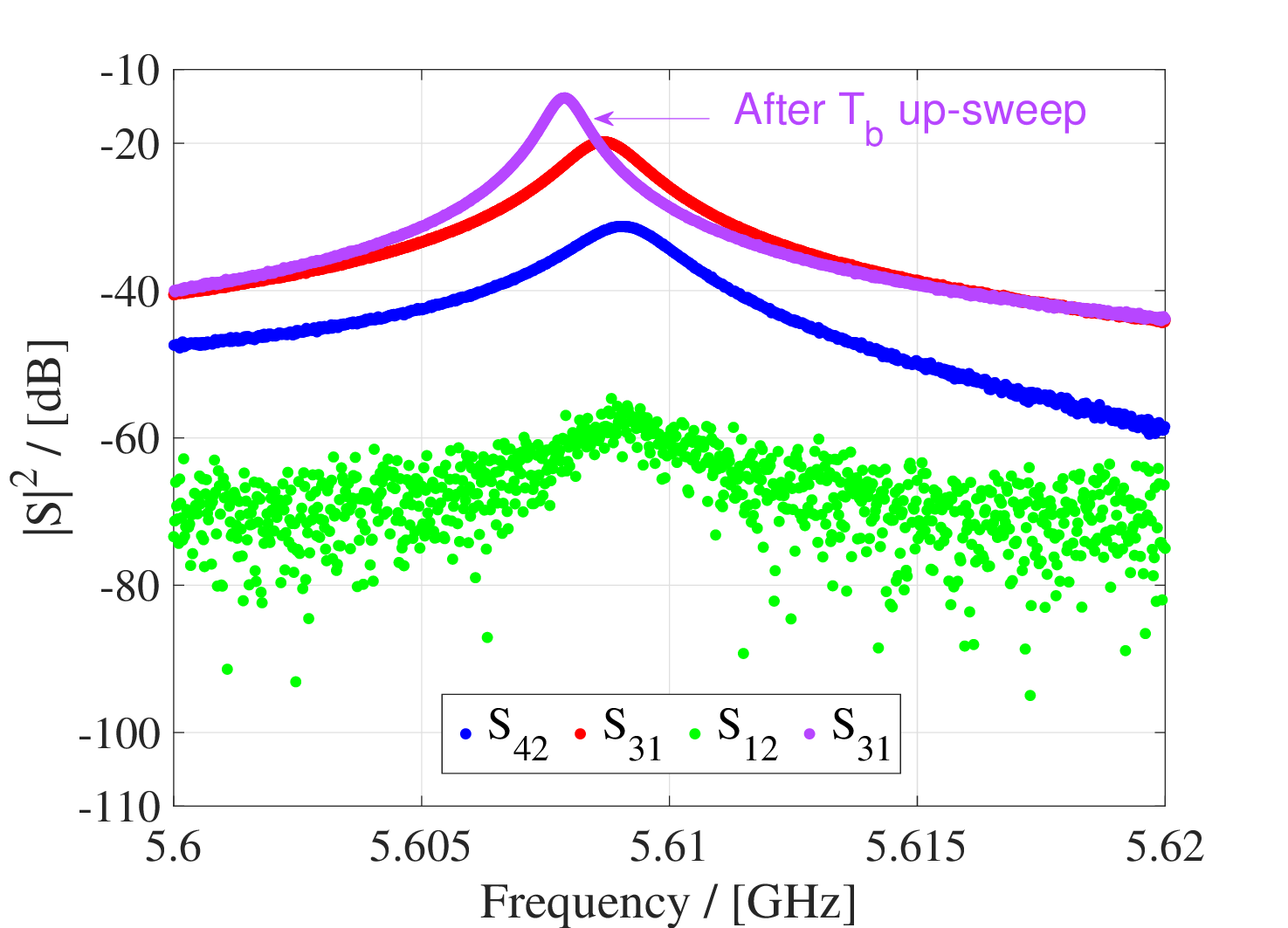}};
    	 \node[inner sep=0, xshift=0cm, yshift=-6.25cm] (image) at (0,0) {\includegraphics[width= .9\linewidth]{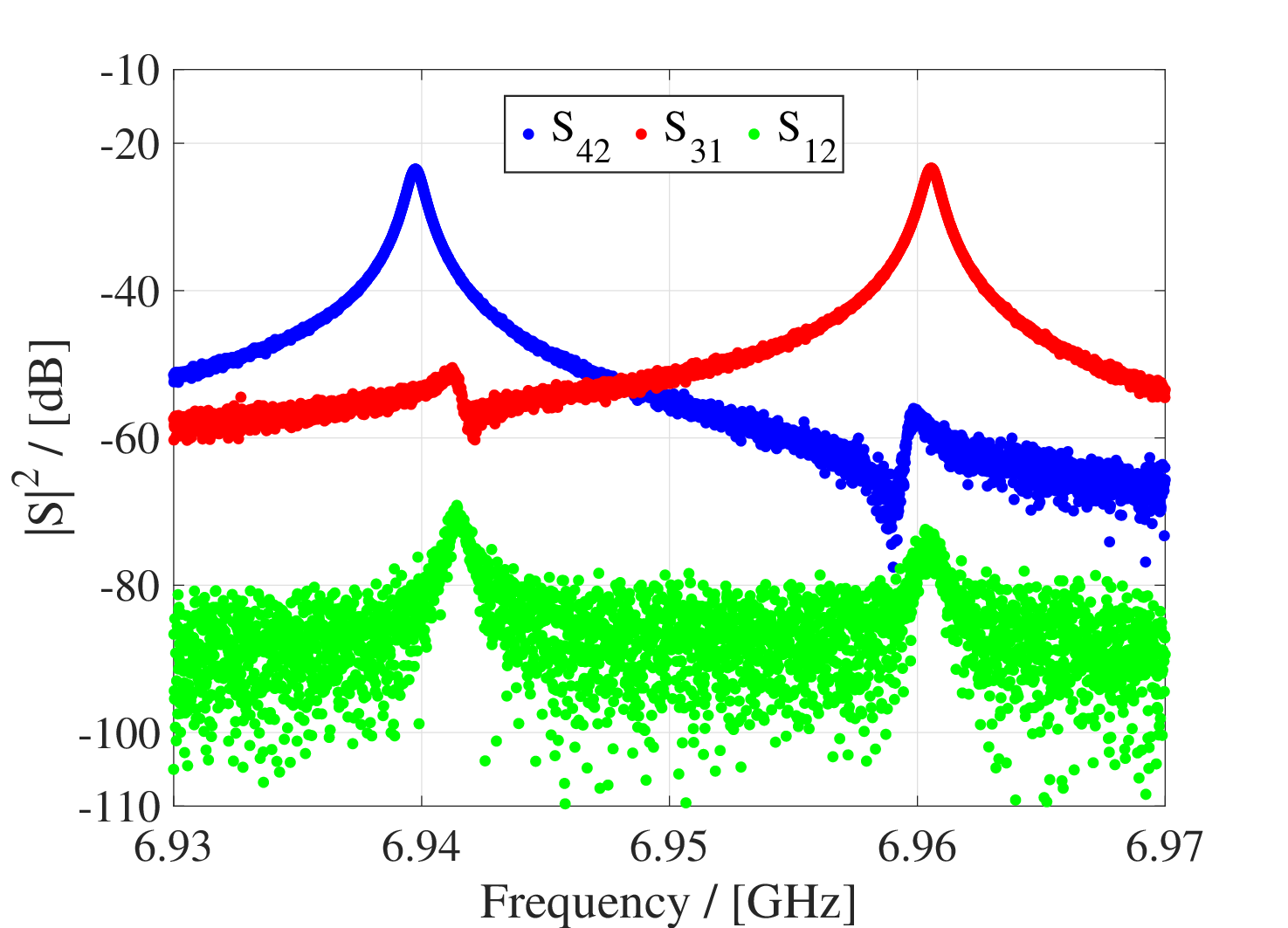}};

    	\node[text=black, scale=1, align=left] at (0.25,-3.25) {(a)};
    	\node[text=black, scale=1, align=left] at (0.25,-9.5) {(b)};
    	
 	 \end{tikzpicture}
\caption[]{\label{fig:S4_cpw_resonance_splitting_measurement} Magnitude of forward transmission between opposite ports (S42 and S31) and transmission between adjacent ports (S12) as a function of frequency at bath temperature $\sim$100mK for CPW ring resonators. (a) Al CPW R1. (b)Nb CPW R1.}
\end{figure}

\begin{figure}[htbp!]
\centering
    \begin{tikzpicture}
        \node[inner sep=0, xshift=0cm, yshift= 0 cm] (image) at (0,0) {\includegraphics[width= .9\linewidth]{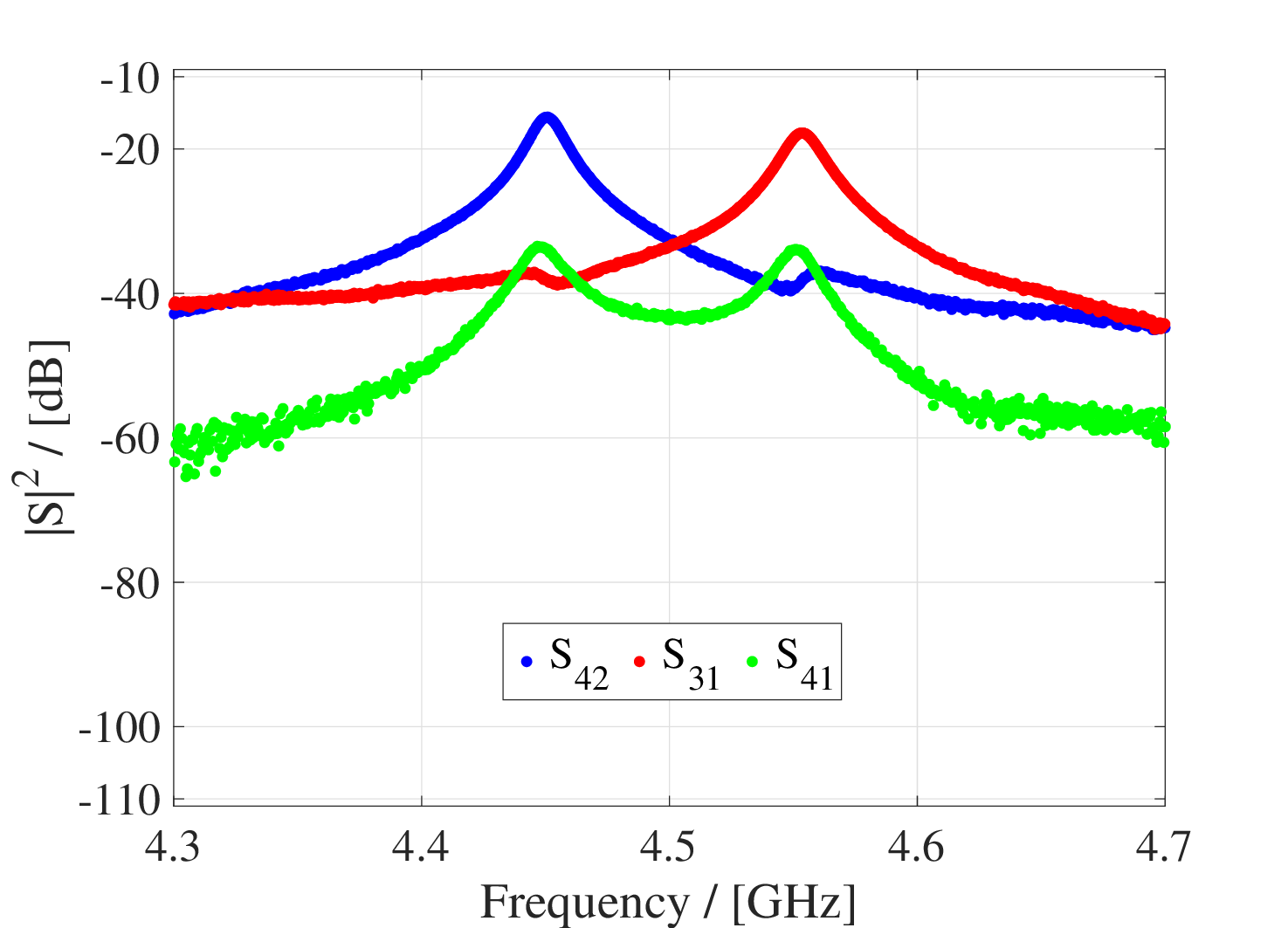}};
        \node[inner sep=0, xshift=0cm, yshift=-6.25cm] (image) at (0,0) {\includegraphics[width=.9 \linewidth]{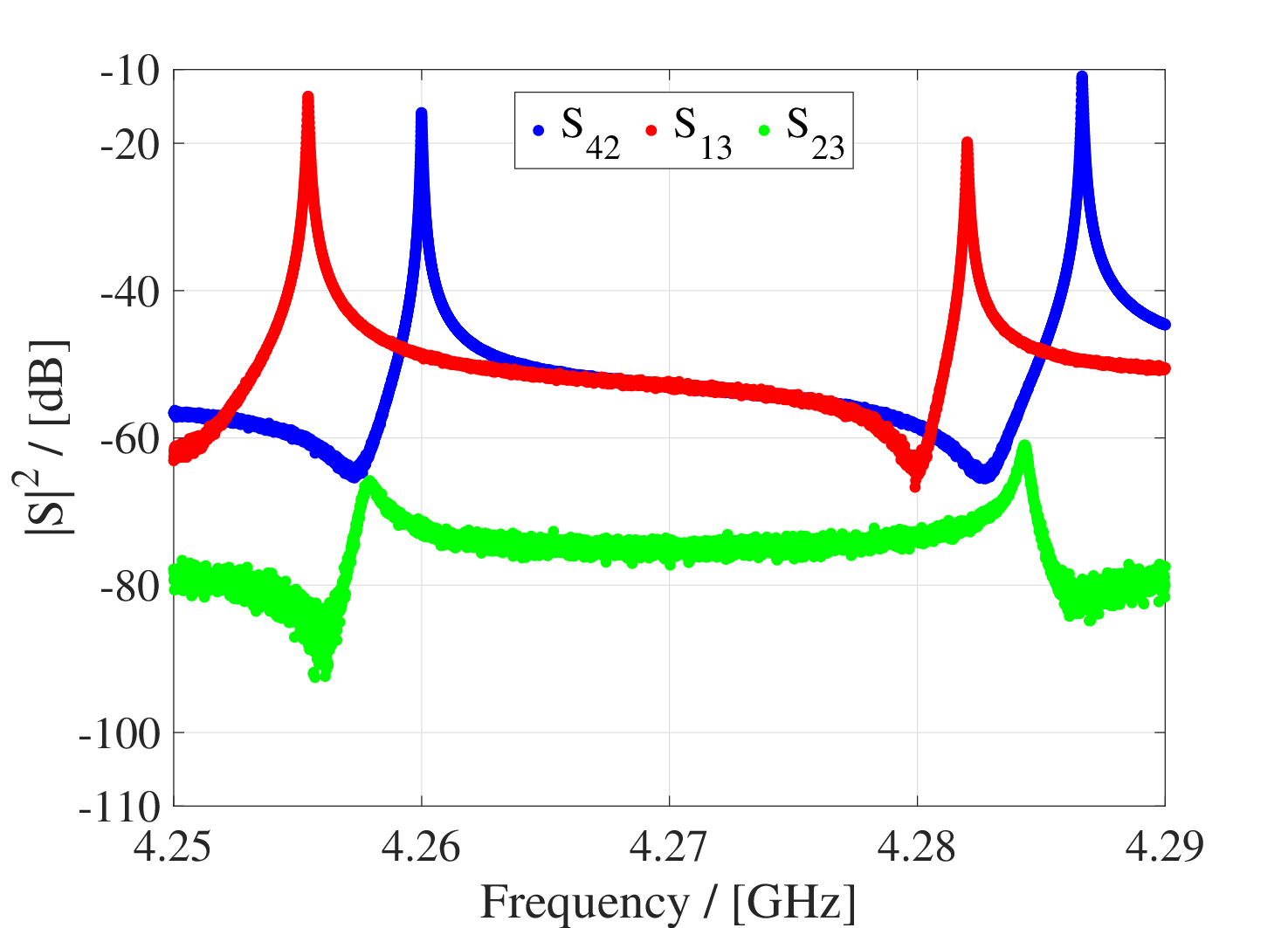}};

        \node[text=black, scale=1, align=left] at (0.25,-3.25) {(a)};
        \node[text=black, scale=1, align=left] at (0.25,-9.5) {(b)};

     \end{tikzpicture}
\caption[]{\label{fig:S4_microstrip_resonance_splitting_measurement} Magnitude of forward transmission between opposite ports (S42, S31, and S13) and transmission between adjacent ports (S41 and S23) as a function of frequency at bath temperature $\sim$100mK for microstrip ring resonators. (a) Al microstrip R1 (b) Nb microstrip R3.}
\end{figure}

The ring resonators were bonded using Al wires to a gold-plated, oxygen-free copper box with SMA connectors, which was attached to the cold stage of an adiabatic demagnetisation refrigerator and tested at $110\pm7\,\mathrm{mK}$. Coaxial cables were used in the cryogenic system, thermally anchored at 70\,K, 4\,K and 1\,K stages, and HEMT amplifier was mounted in the signal chain to amplify the transmission signals. In this study, we focus on two main types of measurements: the through transmission between opposing pairs of ports, i.e. $S_{42}$ and $S_{31}$, and the cross-coupling transmission between adjacent ports. All transmission measurements were taken using a vector network analyser (VNA). The intermediate frequency (IF) bandwidth was set to $1\,\mathrm{kHz}$, and a commercial calibration kit was used to shift the VNA reference plane to the cryostat ports. For all measurements, the VNA sweep power was carefully chosen to avoid saturation or nonlinear effects in the ring resonators or the amplifiers along the signal chain. The full schematic of the ring resonator measurement system is shown in Fig.\ref{fig:S3_measurement_setup_and_ring_device}(a), and measurement results are presented in Fig.\ref{fig:S4_cpw_resonance_splitting_measurement} and \ref{fig:S4_microstrip_resonance_splitting_measurement}.

Fig.\ref{fig:S4_cpw_resonance_splitting_measurement}(a)  Al CPW R1: The $S_{42}$ resonance was measured at 5.60906\,GHz with $Q_r =2760\pm80$, while $S_{31}$ resonance was measured at 5.60866\,GHz with $Q_r =3680\pm70$. The scattering characteristic of this ring resonator is close to the ideal case shown in Fig.\ref{fig:c1_compiled_splitting}(a), where the uniform ring transmission line and strong, near-identical capacitive coupling at the four ports help preserve the symmetry of the ring.
Significantly, when this device underwent a thermal cycle from base temperature to 1.2\,K and back to base temperature, the transmission is enhanced, as indicated by the purple plot, yielding a $Q_r =7200\pm70$. Both the low-Q resonance and the high-Q resonance were stable over hours of measurement. Further, when the device was re-measured in a later set of experiments using the same cryogenic system, similar variation with thermal cycles was observed, both increasing and decreasing the Q-factor. Notably, this behaviour was never observed in any linear superconducting resonators tested over many years and across a wide variety of device types, including those deposited and patterned under the same conditions as the ring resonators. This experimental observation may likely be due to flux trapping, a fundamental phenomenon in superconducting rings influenced by the cool-down history of the cryogenic system \cite{schweitzer1967hysteresis}.

Fig.\ref{fig:S4_cpw_resonance_splitting_measurement}(b) Nb CPW R1: The $S_{42}$ resonance was measured at 6.94\,GHz with $Q_r =9650\pm120$ and a small transmission peak at around 6.96\,GHz.  The $S_{31}$ resonance was measured at 6.96\,GHz with $Q_r =9800\pm50$ and a small resonance feature was measured at around 6.94\,GHz. The transmission in $S_{12}$ indicates approximately 50\,dB of isolation between the two channels.

Fig.\ref{fig:S4_microstrip_resonance_splitting_measurement}(a) Al microstrip R1: The $S_{42}$ resonance was measured at 4.45\,GHz with $Q_r =337\pm7$ and a small resonance feature was measured at around 4.55\,GHz. The $S_{31}$ resonance was measured at about 4.55\,GHz with $Q_r =278\pm5$ and a small resonance feature was measured at around 4.45\,GHz. The double-peak transmission in $S_{41}$ indicates approximately 20\,dB of isolation between two channels. Typically, to ensure low-loss operation of superconducting resonators, the bath temperature $T_b$ needs to be less than 0.1$T_c$, where $T_c$ is the superconducting transition temperature of the material \cite{zmuidzinas2012superconducting}.  Our Al ring resonators exhibited slightly higher losses, likely because the ratio $T_b/T_c$ was only approximately 0.1, rather than being much smaller. As a result, thermal quasiparticle loss limited the total Q-factor.

Fig.\ref{fig:S4_microstrip_resonance_splitting_measurement}(b) Nb microstrip R3: Two-peak resonances of $S_{42}$ (4.255\,GHz and 4.282\,GHz) and $S_{31}$ ( 4.26\,GHz and 4.287\,GHz,) were measured with significant splitting of approximately 30\,MHz. The highest-transmission peaks of $S_{42}$ and $S_{31}$ were charcterised by $Q_r =60100\pm600$ and $Q_r =58000\pm900$ respectively. The frequency splitting is likely due to fabrication inhomogeneities, which introduced impedance perturbations that in turn resulted in significant frequency splitting and mode rotation. Additionally, the peaks measured in the adjacent ports $S_{23}$ are centred between the two peak frequencies of the orthogonal channels ($S_{42}$ and $S_{31}$).

In both Fig.\ref{fig:S4_cpw_resonance_splitting_measurement}(b) and Fig.\ref{fig:S4_microstrip_resonance_splitting_measurement}(a), in addition to the primary resonance peaks, small secondary resonance peaks are observable at the resonance frequencies of the orthogonal channels. The measured double-peak behaviour is in close agreement with the simulated behaviour shown in Fig.\ref{fig:c2_compiled_Z_perturbation}(c). As discussed in the theory section, the origin of the double peak lies in the rotation of the underlying resonance modes relative to the ports, caused by impedance perturbations along the ring’s transmission lines. The same measurement system was previously used to characterise linear resonators, where no frequency splitting or mode rotation was observed. This provides confidence in the reliability of the system. Although Al and Nb ring resonators exhibit high sensitivity to perturbations in the form of frequency splitting and mode rotation, the cross-transmission between adjacent ports remains low compared to the primary transmission peaks, implying that the two modes remain remarkably well isolated from each other, despite perturbations caused by inhomogeneities. This demonstrates that superconducting ring resonators allow independent control of the two lowest-order modes, highlighting their potential for practical applications. In practical ring resonator measurements, the double peaks resulting from frequency splitting and mode rotation help distinguish the ring resonances from other possible resonances (e.g. box mode) within the frequency sweep band.
\section{Conclusions}\label{sec_Conclusion}
We have conducted a systematic analysis of superconducting ring resonators, and experimentally characterised the microwave behaviour of Al and Nb ring resonators in both microstrip and CPW geometries. To the best of our knowledge, in-depth analysis of ring resonators based on superconducting transmission lines appears to be limited in the existing literature.

Our theoretical models closely capture the key features observed in the experimental measurements, providing valuable insight into the operational behaviour of these devices. These resonators, capacitively coupled at four symmetric ports, support two orthogonal lowest-order modes that can be independently controlled. Theoretical analyses show impedance perturbations induce frequency splitting and mode rotation relative to the ports, whereas perturbations in wave speed solely shift the resonance frequency. Scattering parameter measurements confirm that both modes can be selectively excited via opposite ports while maintaining partial isolation between adjacent ports.

As confirmed by our measurements, in practice, high-Q ring resonators will exhibit resonance frequency splitting due to the challenges in achieving extremely high fabrication resolution and precise control over the uniformity of characteristic impedance. Our results, both theoretical as well as experimental, demonstrate that superconducting ring resonators are highly sensitive to  minute impedance perturbations, even of the order of one part in a thousand, due to their high Q-factors, leading to frequency splitting, rotation of resonance modes, and loss of isolation. This high sensitivity makes frequency splitting a useful indicator for detecting the presence of impedance perturbations, while also highlighting the importance of carefully controlling line symmetry and impedance. Our experimental measurements also suggest the presence of trapped flux, an intrinsic feature of superconducting rings that will significantly influence device performance. This effect warrants further investigation to fully understand its impact on device stability and quality factors. 

Future research should explore the performance characteristics and diverse applications of superconducting ring resonators. These devices hold significant promise for potential applications in the forms of both passive components, such as filters, couplers, and isolators, and active elements, such as parametric amplifiers and tuneable resonators. In particular, their application as parametric amplifiers is especially promising, since having two highly-isolated electromagnetic modes coupled to a common quasiparticle system could facilitate easier separation between the pump and signal tones. Furthermore, the flux sensitivity of superconducting rings makes them especially well-suited for integration into quantum circuits, and future studies should explore their flux-tuneable properties and potential applications as particle sensors. Overall, superconducting ring resonators have unique properties and broad applicability, with the potential to become key components in superconducting electronics and quantum technologies.

\bibliographystyle{apsrev4-2}
\bibliography{reference}

\end{document}